\documentclass[9pt, conference]{IEEEtran}
\IEEEoverridecommandlockouts
\usepackage{cite}
\usepackage{amsmath,amssymb,amsfonts}
\usepackage{algorithmic}
\usepackage{graphicx}
\usepackage{textcomp}
\usepackage{xcolor}
\usepackage{array}
\usepackage{booktabs}
\def\BibTeX{{\rm B\kern-.05em{\sc i\kern-.025em b}\kern-.08em
    T\kern-.1667em\lower.7ex\hbox{E}\kern-.125emX}}
\begin{document}

\title{Integrating Pause Information with Word Embeddings in Language Models for Alzheimer's Disease Detection from Spontaneous Speech\\
\thanks{\IEEEauthorrefmark{1}Corresponding author}
\thanks{This work was supported by the National Natural Science Foundation of China under Grant No. 62276153.}
}

\author{\IEEEauthorblockN{Yu Pu, Wei-Qiang Zhang\IEEEauthorrefmark{1}}
\IEEEauthorblockA{
Department of Electronic Engineering, Tsinghua University, Beijing 100084, China \\
\tt puy23@mails.tsinghua.edu.cn, wqzhang@tsinghua.edu.cn}}

\maketitle

\begin{abstract}
Alzheimer's disease (AD) is a progressive neurodegenerative disorder characterized by cognitive decline and memory loss. Early detection of AD is crucial for effective intervention and treatment. In this paper, we propose a novel approach to AD detection from spontaneous speech, which incorporates pause information into language models. Our method involves encoding pause information into embeddings and integrating them into the typical transformer-based language model, enabling it to capture both semantic and temporal features of speech data. We conduct experiments on the Alzheimer's Dementia Recognition through Spontaneous Speech (ADReSS) dataset and its extension, the ADReSSo dataset, comparing our method with existing approaches. Our method achieves an accuracy of 83.1\% in the ADReSSo test set. The results demonstrate the effectiveness of our approach in discriminating between AD patients and healthy individuals, highlighting the potential of pauses as a valuable indicator for AD detection. By leveraging speech analysis as a non-invasive and cost-effective tool for AD detection, our research contributes to early diagnosis and improved management of this debilitating disease. 
\end{abstract}


\section{Introduction}
\label{sec:intro}

Alzheimer's Disease (AD) is a neurodegenerative condition where patients experience aphasia caused by brain lesions. Traditional methods for detecting Alzheimer's disease are time-consuming and costly, imposing a significant burden on families and society  \cite{world2012dementia,qi2023noninvasive}.  Since speech-based detection of Alzheimer's disease offers advantages such as low cost, non-invasiveness, and convenient operation \cite{chen2023raw,chen2023cross,li2024whisper}, it holds great potential for broad application.

The impact of AD on the content of spontaneous speech has led to the exploration of natural language processing techniques for reliable AD detection \cite{qin2021exploiting}. Two common approaches for AD detection from speech include using domain knowledge-based hand-crafted linguistic features and fine-tuning BERT-based sequence classification models \cite{sundar2020raw}. Both feature-based approaches and fine-tuned BERT models have shown promising results in detecting cognitive impairments related to AD, with fine-tuned BERT models demonstrating superior performance in some cases \cite{pappagari2021automatic}. The importance of linguistic features in AD detection has been highlighted, with linguistic information alone achieving comparable or even better performance than models incorporating both acoustic and linguistic features \cite{balagopalan2021comparing}. 

Another crucial indicator for AD detection lies in pauses within speech \cite{pan2021using}. The detection of Alzheimer's disease through speech pauses has gained significant attention in recent literature. Vincze et al. (2020) \cite{Veronika2021silence} proposed an approach for the early detection of mild cognitive impairment (MCI) and mild AD using temporal speech parameters, such as speech rate, number and length of pauses, and pause rates. They discovered that the proportion of pauses within the overall speech duration grew with the advancement of dementia. Pastoriza-Dominguez et al. (2021) \cite{PASTORIZADOMINGUEZ2022107} conducted a study to characterize the probability density distribution of speech pause durations in AD, amnestic mild cognitive impairment (aMCI) patients, and healthy controls (HC). They found that a lognormal distribution explained the pause duration distribution for all groups, with AD patients exhibiting longer pauses and greater variability compared to healthy controls. 

The pauses within sentences are associated with the content, prompting a natural consideration of integrating pause information into language models to bolster AD detection \cite{reilly2011anomia,hoffmann2010temporal}. Yuan et al. (2021) \cite{Yuan2021pause} demonstrated that pauses, disfluencies, and language problems in speech can be modeled using transformer-based pre-trained language models, achieving high accuracy in AD detection. Wang et al. (2023) \cite{Wang2023prompt} investigated the use of prompt-based fine-tuning of pre-trained language models for AD detection, incorporating disfluency features based on hesitation or pause filler token frequencies. Their system achieved a mean detection accuracy of 84.20\% using manual speech transcripts and 82.64\% using ASR speech transcripts on the Alzheimer's Dementia Recognition through Spontaneous Speech (ADReSS) dataset. 

While there have been some attempts to integrate pause information with linguistic features, there has yet to be a method that can deeply incorporate pause information into language models. In this paper, we propose an approach that combines pauses with textual content during the encoding phase of the language model, in order to capture both semantic and temporal features of speech data, thereby enhance its performance in AD detection.

\begin{figure}[htb]
	\begin{minipage}[b]{1.0\linewidth}
		\centering
		\centerline{\includegraphics[width=9cm]{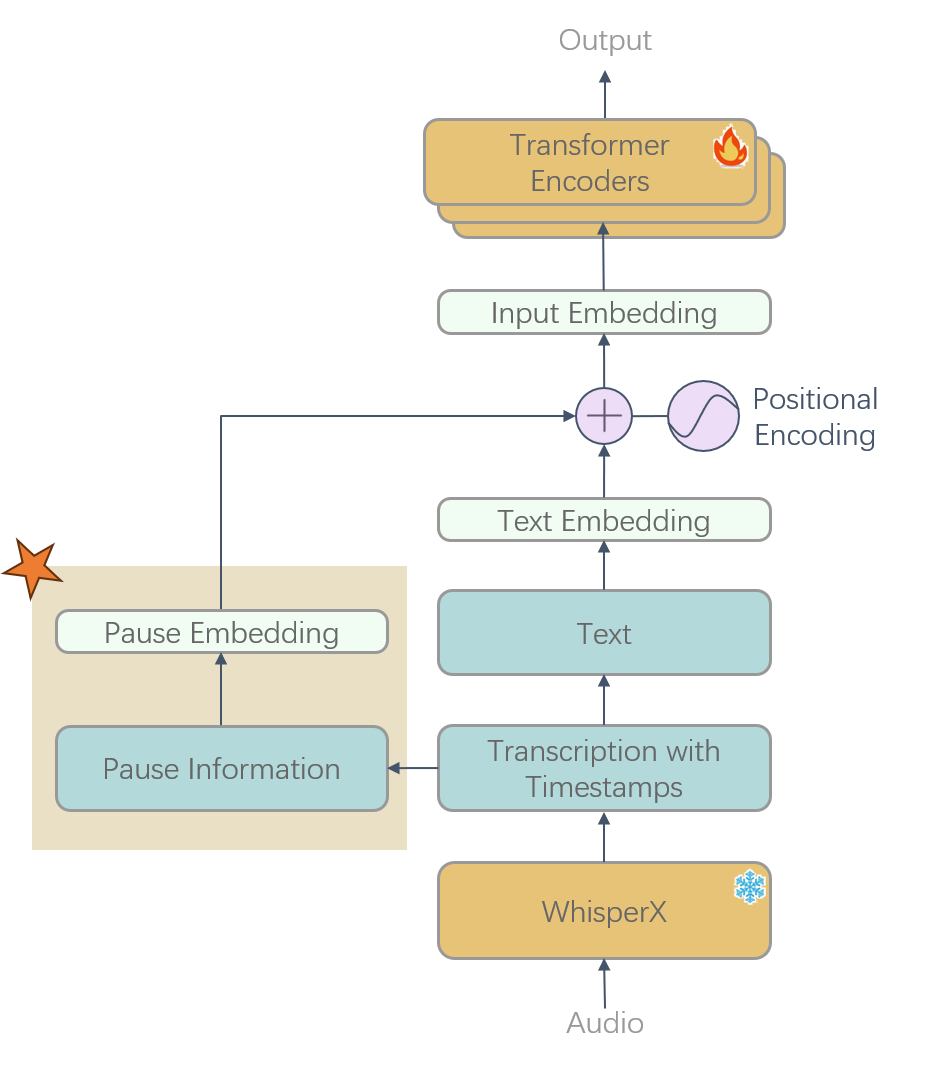}    }
	\end{minipage}
	\caption{General flowgram of our AD detection system. Our main innovation lies in the part marked with a pentagram, which integrates pause information with the language model.}
	\label{fig:Flowgram}
\end{figure}

To validate the effectiveness of our approach, we conduct experiments on two widely used datasets: the ADReSS dataset and its extension, the ADReSSo dataset. We compare our method with existing approaches and demonstrate its superiority in discriminating between AD patients and healthy individuals.

\section{Method}
\label{sec:method}

\subsection{Speech recognition with timestamps} 
\label{ssec:whisperx}
The first step in our methodology involves transcribing speech recordings to text. For this purpose, we utilize WhisperX \cite{bain2023whisperx}, a time-accurate speech recognition system with word-level timestamps. WhisperX addresses challenges related to inaccuracies in timestamps and the lack of word-level timestamps. The system utilizes voice activity detection and forced phoneme alignment to achieve this. By leveraging WhisperX, we obtain accurate transcriptions of spontaneous speech, which serve as the input data for our subsequent analysis. 

\subsection{Pauses and word durations extraction}
\label{ssec:pause extraction}
\begin{figure}[htb]
	\begin{minipage}[b]{1.0\linewidth}
		\centering
		\centerline{\includegraphics[width=9cm]{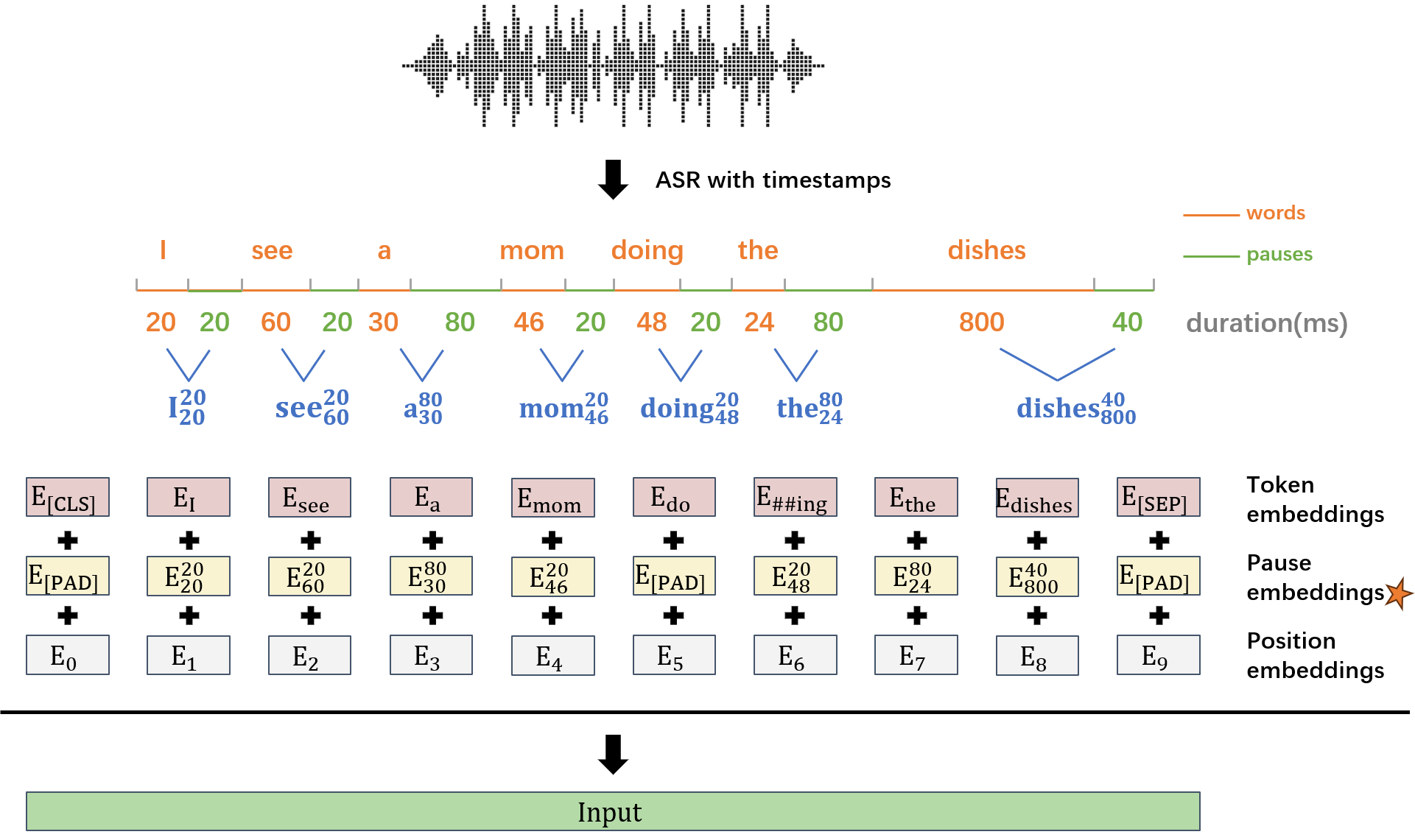}    }
	\end{minipage}
	\caption{The detailed process of embedding calculation.}
	\label{fig:Embeddings}
\end{figure}

To incorporate pause information into the BERT model, we introduce a novel method for encoding temporal features into embeddings. After transcribing spontaneous speech using WhisperX, we extract pauses between consecutive words in the transcribed text. These pauses are then encoded into embeddings using a pause encoder, by which we capture the temporal dynamics of pauses during speech. The detailed process is illustrated in Figure \ref{fig:Embeddings}. We combine the duration of each word with the subsequent pause to form a pause token, where both the duration and pause are constrained within 0-3 seconds and quantized uniformly at 10ms intervals, resulting in a total vocabulary size of 300*300 for pause tokens. 

\subsection{Integration of pause information into language models}
\label{pause integration}
The Bidirectional Encoder Representations from Transformers (BERT) \cite{devlin2018bert} model has demonstrated remarkable performance across various natural language processing tasks, including text classification and language understanding. BERT is pre-trained on large-scale text corpora using a masked language modeling objective, enabling it to capture rich contextual information from input sequences. In our methodology, we leverage the pre-trained BERT model as our base model to capture the semantic information of speech content. 

The encoded pause embedding are integrated into the BERT model architecture, alongside the existing word and position embedding. The specific procedure is also depicted in Figure \ref{fig:Embeddings}. We employ a learnable embedding mapping method to encode both the duration and pause into a 384-dimensional embedding separately. Subsequently, these two embeddings are concatenated in the feature dimension to form a 768-dimensional pause embedding, which is then added to the word embedding and position embedding, thereby integrating pause information with textual information in the BERT model. 

\subsection{Model training and fine-tuning}
\label{model training}
To effectively leverage the rich contextual information present in spontaneous speech data, we adopt a two-stage training approach. In the first stage, we train our learnable pause embedding on the GigaSpeech \cite{chen2021gigaspeech} dataset, a large-scale corpus of transcribed speech data spanning diverse domains and linguistic variations. Training on GigaSpeech enables the model to learn generic features from a broad spectrum of speech samples, thereby facilitating better generalization to downstream tasks.

During training, the model learns to encode both semantic and temporal information from the input speech transcripts, including pause durations encoded as pause embeddings. We employ a masked language modeling objective, similar to the original BERT training procedure, where a random subset of input tokens is masked, and the model is trained to predict these masked tokens based on contextual information from surrounding tokens.

In the second stage, we fine-tune the BERT model on the ADReSSo dataset, a specialized dataset containing speech samples from individuals diagnosed with Alzheimer's disease and healthy controls. Fine-tuning involves optimizing the model parameters specifically for the task of AD detection, using a linear classification head after the BERT model.
 

\section{Experiment}
\label{sec:exp}

\subsection{Dataset}
\subsubsection{GigaSpeech}
\label{gigaspeech}
GigaSpeech \cite{chen2021gigaspeech} is a large-scale speech dataset designed to facilitate research in speech recognition, natural language processing, and related fields. It comprises a diverse collection of speech data sourced from various sources, including broadcast media, audiobooks, podcasts, and conversational recordings. GigaSpeech is notable for its extensive coverage of linguistic diversity, spanning multiple languages, dialects, and genres.

\subsubsection{ADReSS}
In our research, we utilize the Alzheimer's Dementia Recognition through Spontaneous Speech (ADReSS) \cite{luz2021alzheimer} dataset, a benchmark dataset provided by the ADReSS Challenge at INTERSPEECH 2020. The dataset comprises speech recordings and transcripts of spoken picture descriptions elicited using the Cookie Theft picture 
from the Boston Diagnostic Aphasia Exam. It contains speech from 78 non-AD subjects and 78 AD subjects.


\subsubsection{ADReSSo}
 The ADReSSo \cite{luz2021detecting} dataset is an extension of the ADReSS dataset, which served as the foundation for the Alzheimer's Disease Speech Detection Challenge held at the 2021 INTERSPEECH international conference. Comprising distinct training and test subsets, both presented in English, the dataset encompasses 87 instances from individuals diagnosed with Alzheimer's disease, alongside 79 instances from healthy counterparts, culminating in a duration of 5.05 hours. 

\subsection{Experimental Setting}
\label{exp set}
We train our learnable pause embedding on four NVIDIA GeForce RTX 3090 GPUs. During training, we use a learning rate of 1e-5, a batch size of 8, and train the model for 10 epochs. For fine-tuning, we utilize a single NVIDIA GeForce RTX 3090 GPU. We adjust the learning rate to 2e-5 and reduce the batch size to 4 to accommodate the size of the AD dataset. And the BERT model is fine-tuned on the ADReSS and ADReSSo datasets for 200 iterations, respectively.
 
\section{Results}
\label{sec:results}

\subsection{Pause Prediction Task}
In this section, we present the results of our pause prediction experiments using BERT with the pause embedding on both the ADReSS and ADReSSo datasets. The objective of these experiments is to evaluate the model's ability to predict pause durations in transcribed speech data and assess the differences in pause prediction performance between Alzheimer's disease patients and healthy elderly individuals. 

\begin{figure}[htb]
	\begin{minipage}[b]{1.0\linewidth}
		\centering
		\centerline{\includegraphics[width=9cm]{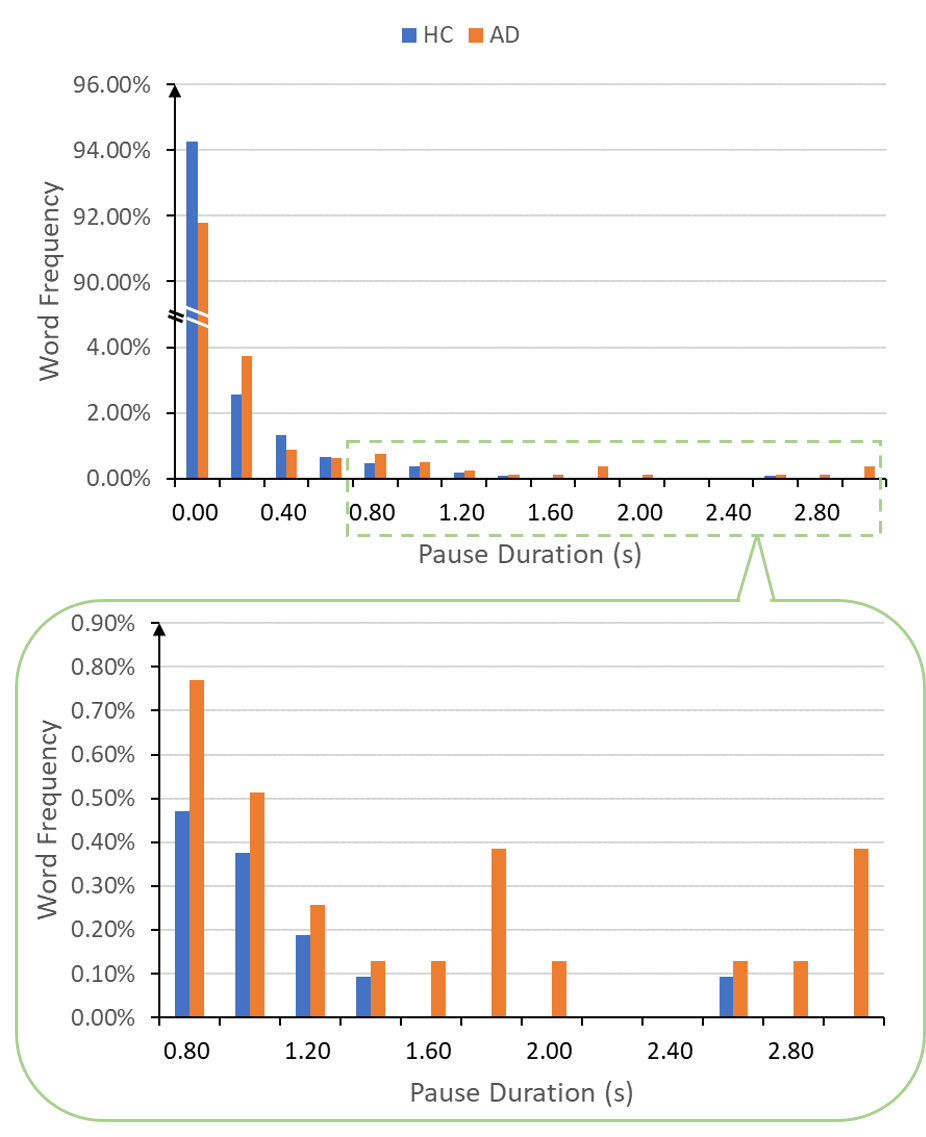}    }
	\end{minipage}
	\caption{The distribution of pause durations in the ADReSSo dataset, with pause duration between 0.8s and 3s amplified.}
	\label{fig:pause distribution}
\end{figure}

Initially, we conduct an analysis of the distribution of pauses on the ADReSSo dataset, as illustrated in Figure \ref{fig:pause distribution}. Most pauses have a duration close to 0, because the spontaneous speech flows smoothly and continuously for most of the time. But after amplifying the distribution of pauses exceeding 0.8 seconds, we can find a propensity among AD patients to demonstrate increased frequency and duration of pauses in their speech patterns. This observation suggests the potential utility of pauses as indicators for evaluating AD. Furthermore, this analysis serves to validate the efficacy of WhisperX in accurately extracting pauses from speech. 

To conduct the pause prediction experiment, we systematically mask the pause tokens corresponding to each word in the transcribed text and task the model with predicting these masked tokens. We then calculate the Root Mean Square Error (RMSE) of the model's predictions across all words in the transcribed text. 

\begin{figure}[htb]
\centering
    \begin{minipage}[b]{0.45\linewidth}
        \centering
        \centerline{\includegraphics[width=4.5cm]{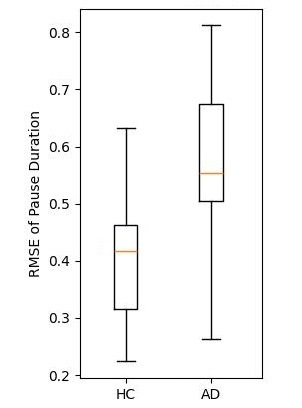}    }
        \centerline{(a) ADReSS}\medskip
    \end{minipage}
    \begin{minipage}[b]{0.45\linewidth}
        \centering
        \centerline{\includegraphics[width=4.5cm]{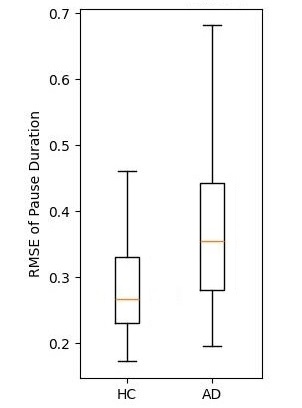}}
        \centerline{(b) ADReSSo}\medskip
    \end{minipage}
    \caption{Results of the pause prediction task. }
    \label{fig:pause res}
\end{figure}

The results of the pause prediction experiment are visualized in Figure \ref{fig:pause res}. In both the ADReSS and ADReSSo datasets, we observe that the mean, minimum, and maximum RMSE values for AD patients are consistently higher than those for healthy elderly individuals. This indicates that the model encounters greater difficulty in predicting pause durations for AD patients compared to healthy controls. Since the learnable pause embedding is trained on a dataset consisting of speech data from healthy individuals, this result suggests that there are significant differences in pause durations between AD patients and healthy individuals. These findings validate our initial hypothesis and demonstrate that our model can effectively extract pause information from speech data. 

The observed differences in pause prediction performance between AD patients and healthy elderly individuals underscore the potential utility of pause information as a discriminative feature for AD detection. The model's difficulty in predicting pause durations for AD patients suggests that AD-related speech patterns manifest in distinct temporal dynamics, which may serve as valuable biomarkers for early detection and monitoring of the disease. Furthermore, these results highlight the effectiveness of our model in capturing subtle acoustic cues related to pauses in spontaneous speech, further supporting its utility in AD detection tasks. 

\subsection{Alzheimer's Detection Task}

In this section, we present the results of our Alzheimer's disease detection experiments conducted on both the ADReSS and ADReSSo datasets, focusing on the binary classification task of distinguishing between AD patients and healthy individuals. To provide a comparison with previous research that utilized pause information to enhance language models, we first replicate the method proposed by Yuan et al. on the ADReSS dataset. We achieve results consistent with those reported in their paper. Yuan's method utilized manually transcribed text provided with the ADReSS dataset. To align with our approach, we apply the replicated method to automatically transcribed text obtained using the WhisperX speech recognition model. The results are shown in Table \ref{tab:ADReSS}. 
\begin{table}[htb]
  \caption{Alzheimer's detection tasks on ADReSS.}
  \vspace{0.2cm}
  \centering
  \scalebox{1.0}{ 
  \begin{tabular}{lcccc} \hline 
    \textbf{Model}  & \textbf{Acc(\%)}& \textbf{Pre(\%)}& \textbf{Rec(\%)}&\textbf{F1(\%)}\\ \hline 
 BERT& 70.8& 75.0& 62.5&68.2\\
    BERT3p \cite{yuan2020disfluencies}& 77.1& 88.2& 62.5&73.2\\
    Ours& 81.2& 82.6& 79.2&80.9\\ \hline
  \end{tabular}
  }
\label{tab:ADReSS}
\end{table}

In Table \ref{tab:ADReSS}, BERT3p refers to the method proposed by Yuan et al. in the 2020 ADReSS Challenge. From the results, we can observe the following:
\begin{enumerate}
    \item Incorporating pause information into the BERT model resulted in a significant improvement in both accuracy and F1 score, indicating a positive impact of pause information on the AD detection capability of the language model.
    \item Compared to BERT3p, our approach achieves higher accuracy and F1 score, suggesting that our method of encoding and integrating pause information into the language model is more effective.
\end{enumerate}
These results validate the efficacy of our approach in leveraging pause information to enhance AD detection performance of language models.

We also present the results of our Alzheimer's disease detection experiments conducted on the ADReSSo dataset. We compare our method with other approaches, and the results are shown in Table \ref{tab:ADReSSo}. 

\begin{table}[htb]
  \caption{Alzheimer's detection tasks on ADReSSo.}
  \vspace{0.2cm}
  \centering
  \scalebox{0.9}{ 
  \begin{tabular}{lcccc} \hline 
    \textbf{Model}  & \textbf{Acc(\%)}& \textbf{Pre(\%)}& \textbf{Rec(\%)}&\textbf{F1(\%)}\\ \hline 
 BERT& 80.3& 95.6& 62.8&75.9\\
    BERT3p \cite{yuan2020disfluencies}& 74.6& 74.3& 74.3&74.3\\
 DisFl+ERNIE \cite{qiao2021alzheimer}& 79.0& 79.0& 78.6&79.0\\
    Ours& 83.1& 92.6& 71.4&80.6\\ \hline
  \end{tabular}
  }
\label{tab:ADReSSo}
\end{table}

From Table \ref{tab:ADReSSo}, we can still observe the positive impact of incorporating pause information on the AD detection capability of the BERT model. Furthermore, we compare our method with the replicated BERT3p and DisFl+ERNIE, a method proposed by Qiao et al. in the 2021 ADReSSo Challenge. Our method achieves the highest accuracy and F1 score among approaches using pause information, indicating that our approach of encoding and integrating pause information into the language model is one of the most effective methods for using pause information in AD detection. 

\subsection{Ablation Study}

To further understand the contributions of the duration and pause embedding in our proposed model, we conduct ablation experiments. Specifically, we evaluate the performance of our model on the ADReSSo dataset by removing either the duration embedding or the pause embedding. The results of these experiments are presented in Table \ref{tab:ablation}. 

\begin{table}[htb]
  \caption{Ablation experiments on ADReSSo.}
  \label{tab:ablation}
  \vspace{0.2cm}
  \centering
  \scalebox{1.0}{ 
  \begin{tabular}{lcccc} \hline 
    \textbf{Model}  & \textbf{Acc(\%)}& \textbf{Pre(\%)}& \textbf{Rec(\%)}&\textbf{F1(\%)}\\ \hline 
 Ours& 83.1& 92.6& 71.4&80.6\\
    w/o duration& 80.3& 92.0& 65.7&76.7\\
 w/o pause& 80.3& 88.9& 68.6&77.4\\ \hline
  \end{tabular}
  }
\end{table}

From Table \ref{tab:ablation}, it is evident that the removal of either the duration or pause embedding leads to a performance drop. When duration embedding is excluded, the accuracy decreases to 80.3\%, and the F1 score drops to 76.7\%. Similarly, excluding pause embedding results in an accuracy of 80.3\% and an F1 score of 77.4\%. These declines in performance suggest that both duration and pause embeddings contribute to the model's ability to accurately detect Alzheimer's disease.

The superior performance when both embeddings are included can be attributed to the potential informational synergy between word duration and pauses. In speech, the duration of words and the pauses between them are likely to be correlated, and mismatches in these patterns may indicate the presence of Alzheimer's disease. By leveraging both types of information, our model can more effectively capture the subtle temporal dynamics of speech associated with AD, leading to improved detection performance.



 


\section{Conclusions}
\label{sec:discussion}

Our study focuses on leveraging pause information in spontaneous speech for Alzheimer's disease detection. The incorporation of pause information into language models capitalizes on subtle temporal cues present in spontaneous speech. Our experiments demonstrate that integrating pause information into the BERT model significantly improved its ability to discriminate between AD patients and healthy individuals. This underscores the importance of considering temporal dynamics, such as pauses, in the analysis of speech data for detecting neurodegenerative disorders like AD.

The observed differences in pause prediction performance between AD patients and healthy controls provide valuable insights into AD-related speech patterns. The greater difficulty encountered by the model in predicting pause durations for AD patients suggests that AD-related speech alterations manifest in distinct temporal dynamics. These findings highlight the potential of pause information as a sensitive biomarker for early detection and monitoring of AD.

We compare our method with existing approaches, including replicating a method proposed by Yuan et al. and evaluating performance on the ADReSSo dataset. The results demonstrate the superiority of our approach in leveraging pause information for AD detection, achieving higher accuracy and F1 score compared to other methods. This highlights the effectiveness of our approach in encoding and integrating pause information into language models.

Future research directions include further exploration of temporal features beyond pauses, such as speech rate, rhythm, and intonation, to enhance the discriminatory power of language models for AD detection. Additionally, longitudinal studies are warranted to assess the utility of pause information in tracking disease progression and response to treatment over time. Moreover, investigating the generalizability of our approach to other neurodegenerative disorders and linguistic contexts can broaden its applicability in diverse clinical settings.

In conclusion, our study highlights the potential of pause information as a valuable resource for enhancing AD detection using language models. By incorporating subtle temporal cues into the analysis of spontaneous speech, we pave the way for more accurate, non-invasive, and early detection of AD, ultimately contributing to improved patient outcomes and quality of life.

\bibliographystyle{IEEEbib}
\bibliography{strings,refs}

\vspace{12pt}

\end{document}